# Perturbative and non-perturbative diffraction at HERA

***Marcella Capua[1] on behalf of the H1 and ZEUS Collaborations***

*(1) Department of physics, Università della Calabria and INFN, Arcavacata di Rende (CS) 87036, Italy, capua@cs.infn.it*

**Abstract**

We review recent results on diffraction at HERA and discuss their interpretation in the perturbative and non-perturbative regimes.

## Introduction

The renewed interest in diffractive physics, born in the 60's, is mainly due to the significant results obtained at the *ep* collider HERA (for a review see [1]). Diffractive events at HERA are possible because the electron radiates a virtual photon that can fluctuate into a quark-anti-quark pair that interacts with the proton. For these events the target proton emerges in the final state with an energy approximately equal to that of the incoming proton beam; the colourless exchange involved in the diffractive interaction being called Pomeron. In the final state, together with the scattered electron, there is a proton and an hadronic system X (or a resonance) with a large gap in rapidity between them.

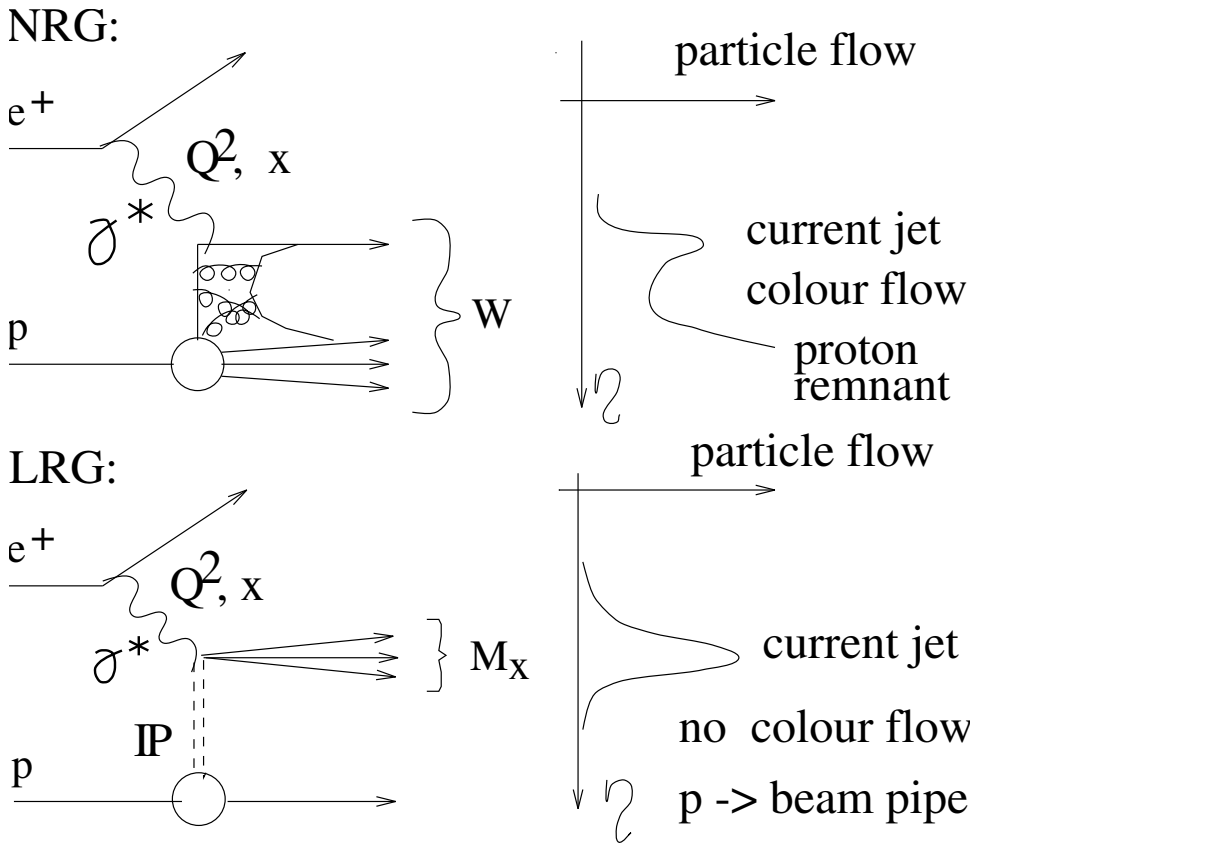

***Fig. 1**: Non-diffractive diagram (top) and inclusive diffractive diagram (bottom) at HERA.*

The upper part of Fig. 1 presents, as an example, the diagram of a standard deep inelastic (DIS) event at HERA with a photon virtuality $Q^2$ and, at the bottom, a typical inclusive diffractive event, $\gamma^* p \rightarrow Xp'$, where the lack of colour flow between the final proton and hadronic system X is visible.

In addition to the standard DIS variables, the following variables are relevant: the diffractive variable $t$, defined as the four-momentum squared transferred at the proton vertex; $x_{IP}$, the longitudinal momentum fraction of the colourless exchange with respect to the incoming proton; and $\beta$, the longitudinal momentum fraction of the struck quark with respect to the colourless exchange. The latter two variables are connected to the Bjorken scaling variable $x$ by the relation $x = \beta \cdot x_{IP}$.

Diffractive interactions can be described in the framework of Regge phenomenology[2] in terms of the exchange of a specific Pomeron trajectory which has the vacuum quantum numbers. In QCD, Pomeron exchange is described, at the leading order, by the exchange of two gluons with the vacuum quantum numbers. The hard scales available at HERA make the diffractive processes treatable within perturbative QCD and this is a subject of intense theoretical and experimental research.

Experimentally, a diffractive event requires no hadronic activity in the direction of the proton flight, as the proton remains intact in the diffractive process. To select diffractive events, both the H1 and the ZEUS experiments adopted three different methods:

- the first method requires a proton detected from a dedicated spectrometer (*proton-tag method*) placed in the forward proton direction (LPS for ZEUS or FPS/VFPS for H1) with a fraction of the initial proton momentum, $x_L$, greater than 97%. With this method the selection is completely free from contamination of events where the proton dissociates into a low mass system Y, $ep \rightarrow eXY$. This is the cleanest method but is affected by a lack of statistics.

- The second method requires a large gap in rapidity (*LRG method*) between the hadronic system X measured in the central detector and the scattered proton direction. This method, rich in statistics, is affected by relevant contamination from proton dissociative events.

- The third method identifies the diffractive and non diffractive contributions on the mass of the system $X$ ($M_X$ *method*) base of their characteristic dependence on $M_X$.

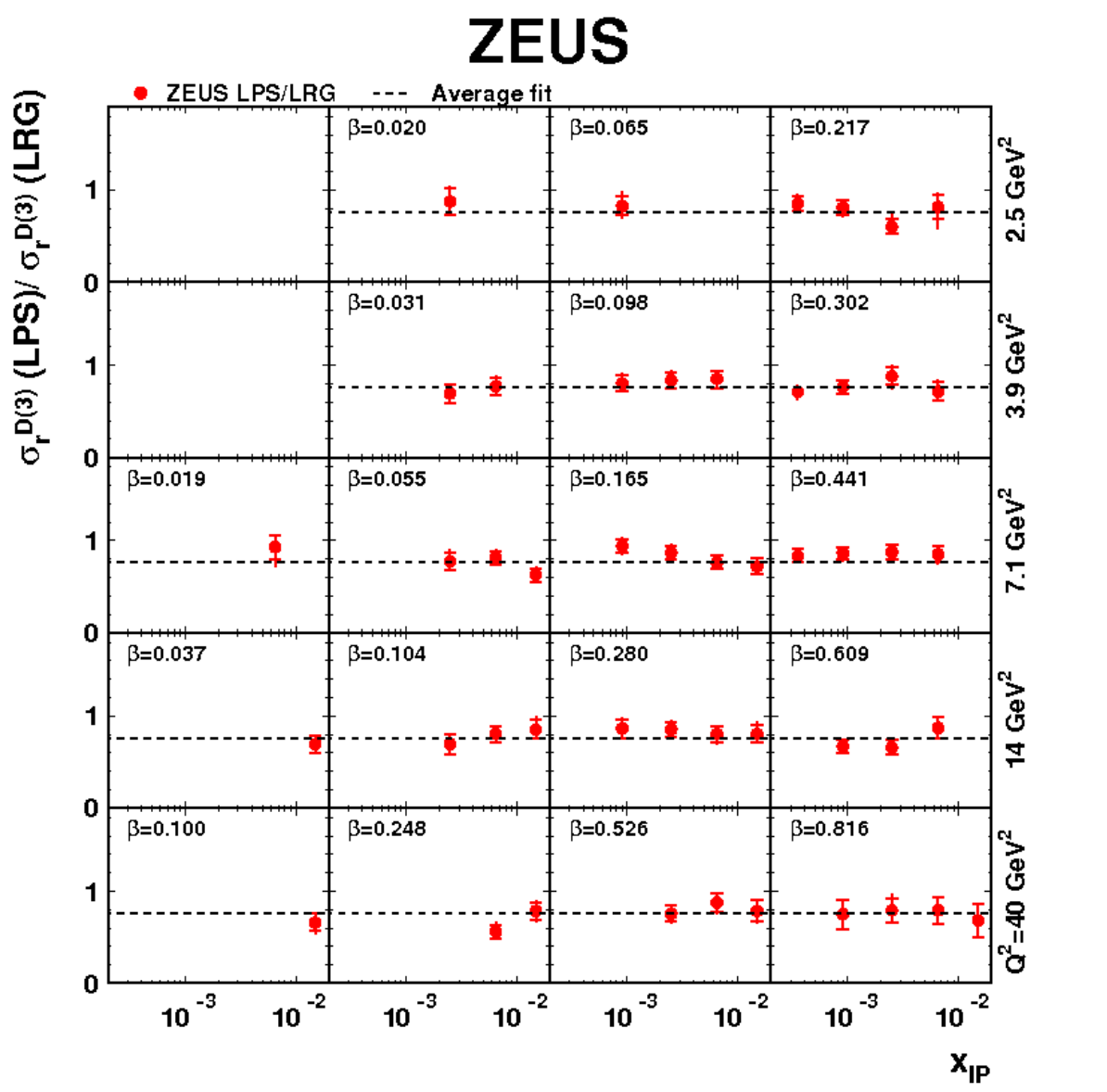

***Fig. 2**: The ratio of the ZEUS LPS measurement to the ZEUS LRG measurement before subtraction of proton dissociative background. The lines represent the average value of this ratio.*

The three methods are complementary with respect to the kinematic coverage and differ substantially in their dominant sources of systematic uncertainty. The comparison between the first two methods represents also the technique adopted to estimate the proton dissociative background selected by LRG method.

Although a detailed discussion of the measurements will be presented in the next section, we anticipate here in Fig. 2 the ZEUS[3] result that shows the ratio of the inclusive diffractive cross section obtained with the LRG method to that with the proton-tag as a function of $Q^2$ and $\beta$. The dotted line shows the average value of the ratio obtained $0.76\pm0.01(\text{stat.})^{+0.03}_{-0.02}(\text{syst.})^{+0.08}_{-0.05}(\text{norm.})$ and indicates a mean contamination of proton dissociation in the LRG measurements of about 24%. This result seems to be independent on $Q^2$, $x_{IP}$ or $\beta$ dependence between the two methods.

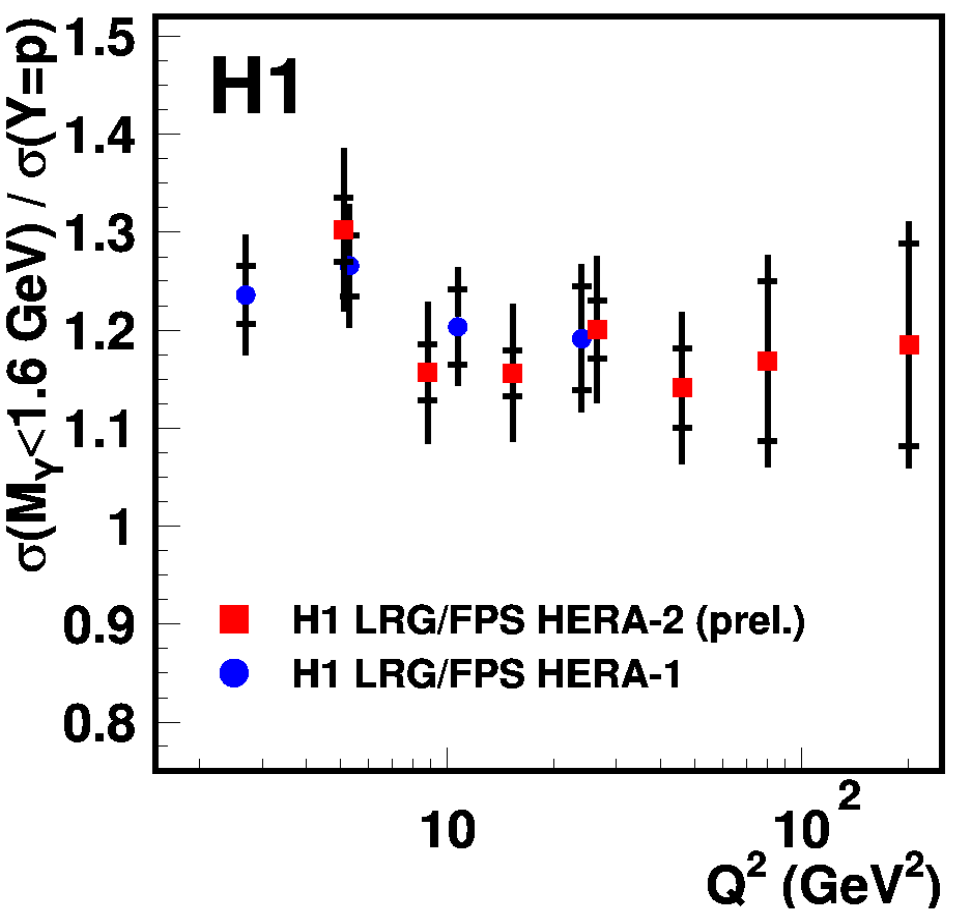

***Fig. 3**: The ratio of the H1 LRG measurement to the H1 FPS measurement, as a function of $Q^2$ before subtraction of proton dissociative background.*

Preliminary[4] and published[5] measurements by H1 result in a consistent value for the ratio $1.23\pm0.03(\text{stat.})\pm0.16(\text{syst.})$ as shown in Fig. 3, where the ratio between the cross sections obtained with the LRG and the proton tagger is presented.

## Inclusive results and DPDFs at HERA

Similarly to inclusive DIS, cross section measurements for the reaction $ep \rightarrow eXp'$ are conventionally expressed in terms of the reduced diffractive cross section, $\sigma_r^{D(4)}$, which is related to the measured four-fold differential cross section by

$$\frac{d^4\sigma}{d\beta\, dQ^2 dx_{IP} dt} = \frac{4\pi\alpha^2}{\beta Q^4}\left[1 - y + \frac{y^2}{2}\right]\sigma_r^{D(4)}(\beta, Q^2, x_{IP}, t)$$

The reduced cross section in turn is related to the diffractive structure functions $F_2^{D(4)}$ and $F_L^{D(4)}$ by

$$\sigma_r^{D(4)}(\beta, Q^2, x_{IP}, t) = F_2^{D(4)} - \frac{y^2}{2(1 - y + y^2/2)} F_L^{D(4)}$$

The reduced cross section corresponds to the diffractive structure function $F_2^{D(4)}$ with a good approximation if the inelasticity $y$ is sufficiently small.

The cross section $\sigma_r^{D(3)}$ and structure functions $F_2^{D(3)}$ and $F_L^{D(3)}$ are obtained from the previous formulas by integration over $t$.

The H1 Collaboration presented this year a new set of inclusive measurements[4,6] based on their entire data sample, obtained with the LRG and VFPS methods, Fig. 4 summarize all the recent results available and shows the extended kinematic coverage.

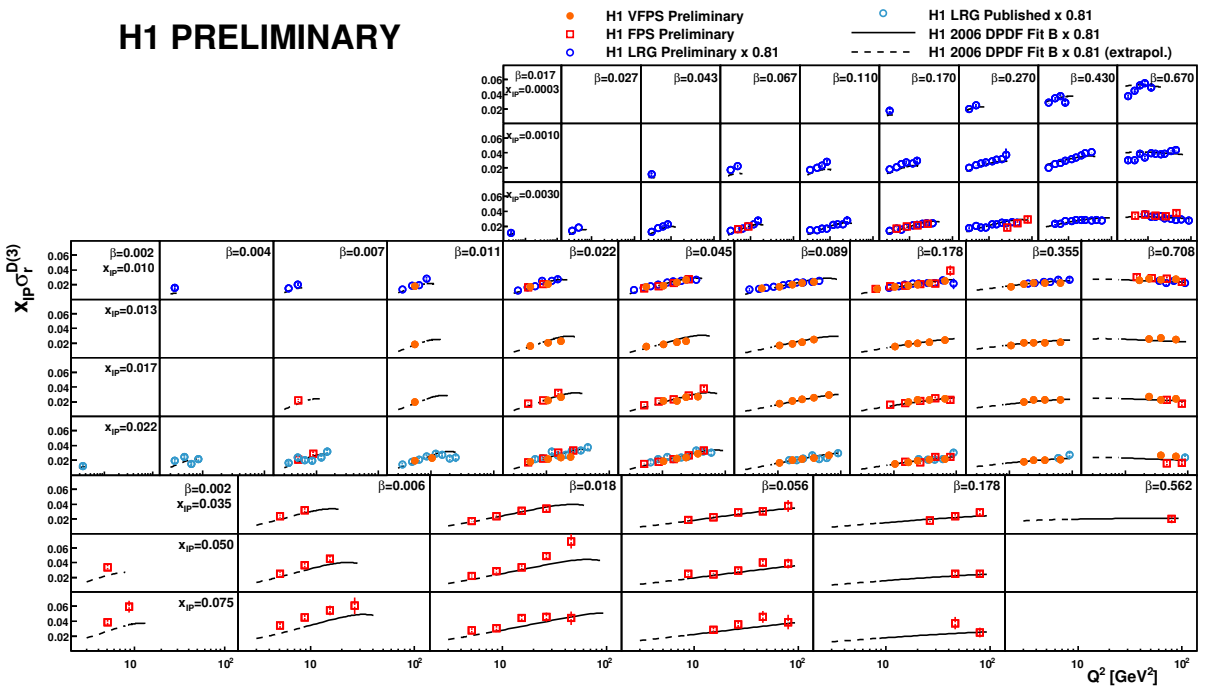

***Fig. 4:*** *Compilation of the* $x_{IP}\cdot\sigma_r^{D(3)}$ *measurements as a function of* $Q^2$ *obtained by H1 with the VFPS and LRG method data and compared with published results[5].*

The recent LRG results from H1 and ZEUS are compatible in most of the kinematic region covered by the measurements and are shown in. Fig. 5.

Work is in progress on the combination of the diffractive cross sections measured by H1 and ZEUS, and this should bring a significant reduction of the systematic uncertainties affecting the results. A first step in this direction is shown in Fig. 6.

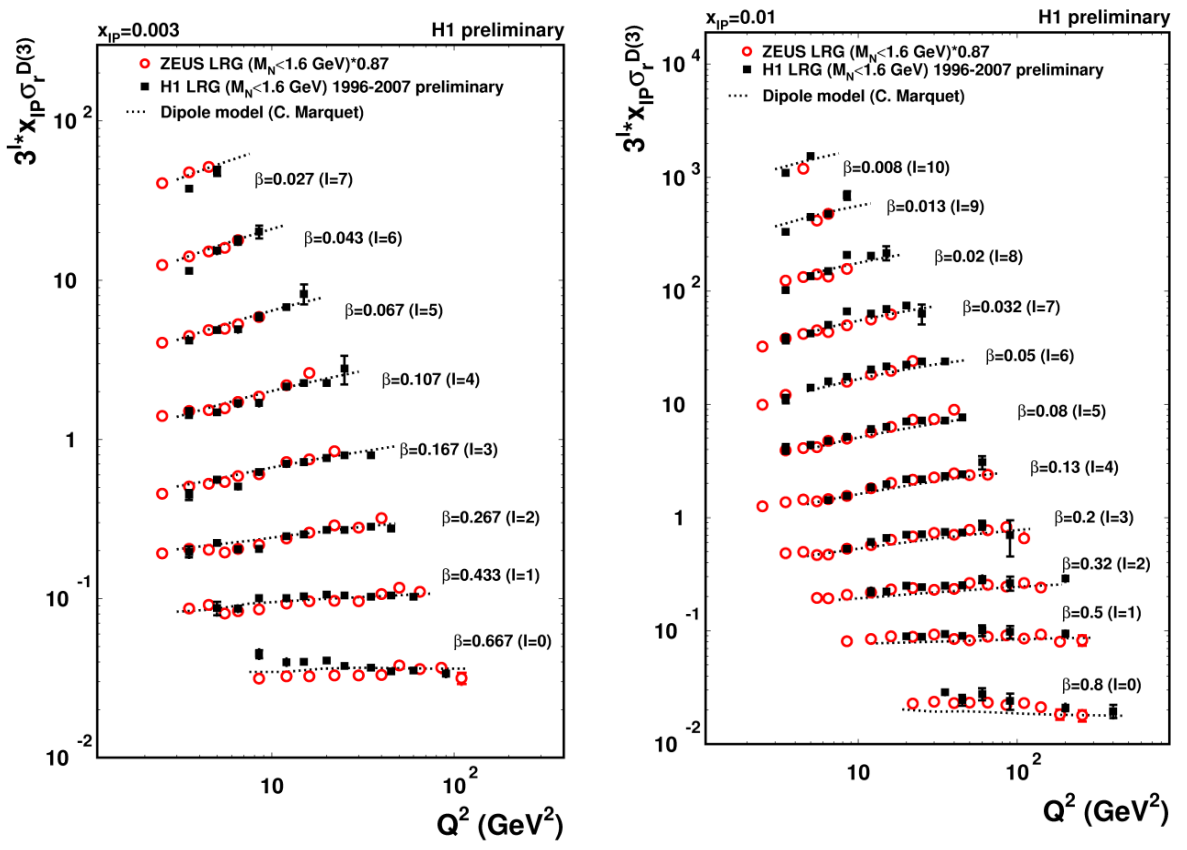

***Fig. 5:*** *Comparison between the H1 and ZEUS LRG measurements after correcting both data sets to* $M_Y < 1.6$ *GeV and applying a further scale factor of 0.87 (corresponding to the average normalization difference) to the ZEUS data for two different* $x_{IP}$ *values[6].*

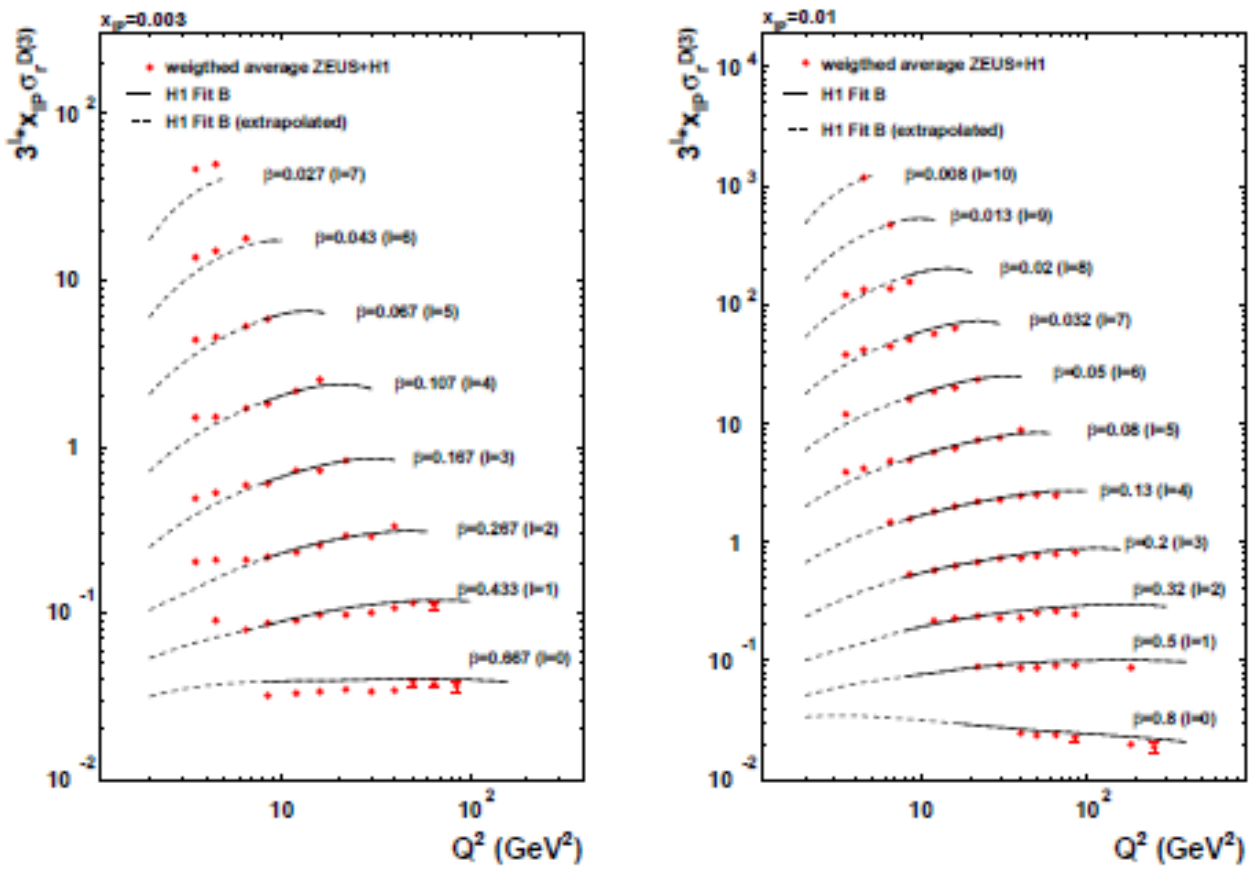

***Fig. 6:*** *First combination of the H1 and ZEUS LRG results presented in [7].*

Inclusive diffraction, according to the factorisation theorem [8], can be described as a convolution of a hard scattering cross section and diffractive parton density functions (DPDFs) which describe the distribution of partons in the proton in a process containing a fast proton in the final state:

$$\sigma(\gamma^* p \rightarrow Xp) \approx f_{IP}(x_{IP}, t) \times f_i(z, Q^2) \otimes \hat{\sigma}_{\gamma^* i}(z, Q^2)$$

In this formula the proton-vertex factorization hypothesis has been assumed and Figs. 2 and 3 show, as a confirmation of the approximate validity of this ansatz, the independent behaviour of $(x_{IP},t)$ from $(Q^2,z)$.

In this way DPDFs are parameterized as a function of the hard scale, $Q^2$, and the longitudinal momentum fraction, $z$, of the parton entering the hard sub-process; $z$ is a generalisation of the $\beta$ variable already presented.

The extraction of the DPDFs has been performed by both the H1 and ZEUS collaborations and the DPDFs have been used successfully to describe other diffractive DIS data. Jet data have also been used in the fits, together with inclusive data, to improve the extraction of the gluon density. Details of the analyses are available in [5,9,10,11].

The predictive power of the HERA DPDFs has been widely presented in recent conferences. It is expected that a QCD analysis based on the H1 and ZEUS combined results should allow the extraction of the most precise DPDFs to date. These could be effectively used for predictions of the diffractive production of the Higgs boson at the LHC.

We also presented a recent update[12] of the first measurement of the longitudinal diffractive structure function $F_L^{D(3)}$ obtained by H1. $F_L^{D(3)}$ is related to the diffractive gluon density and can provide a test of diffractive factorisation and of the role of gluons, complementary to jet and charm data. Figure 7 shows the H1 results as a function of $\beta$ for two different values of $Q^2$ and at a fixed value of $x_{IP}$. The results are compared and consistent with the predictions based on the H1 NLO QCD fit published for $Q^2$ >8.5 GeV$^2$ and its extrapolation down to $Q^2$=4 GeV$^2$.

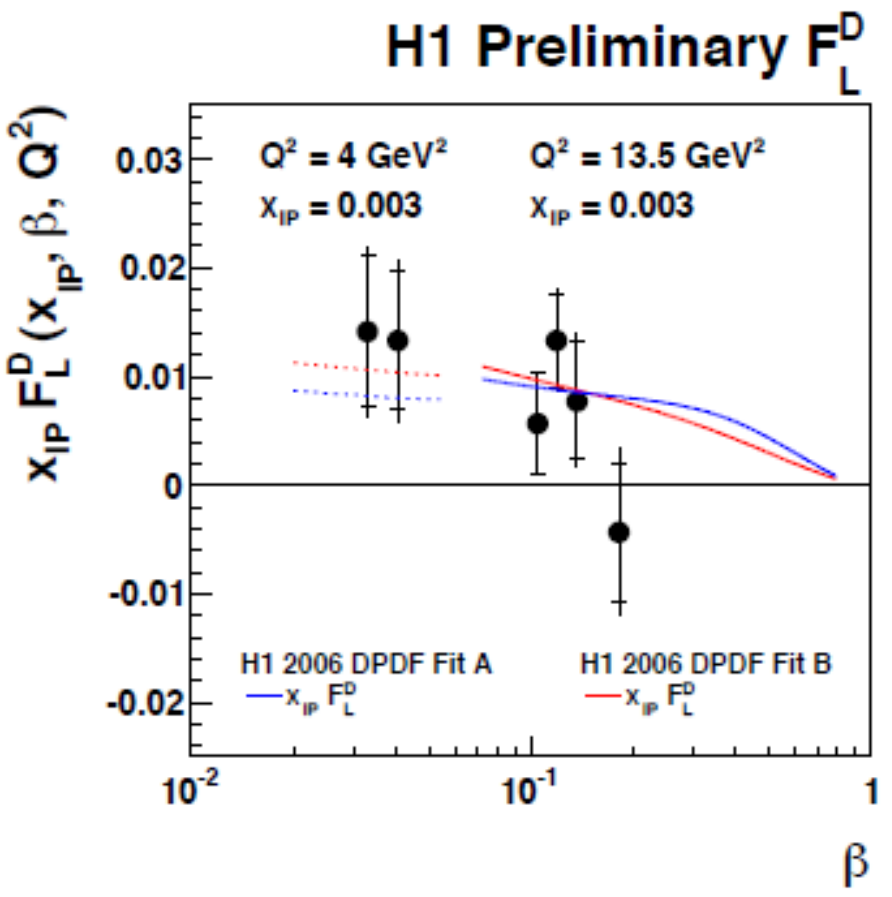

***Fig. 7:*** *Diffractive longitudinal strucuture function multiplied by $x_{IP}$ as a function of $\beta$.*

## Leading Baryon measurements at HERA

The leading baryon production in DIS is interesting because it provides a testing ground for the theory of strong interactions in the soft regime. H1 and ZEUS extensively studied the production of leading protons and leading neutrons in DIS and photoproduction (PHP) using dedicated detectors (recent publications are [13,14]). The dynamical mechanisms for their production are not completely understood. Leading neutron (LN) production occurs through the exchange of isovector particles, notably the $\pi^+$ meson. For leading proton (LP) production isoscalar exchange also contributes, including diffraction mediated by Pomeron exchange. As shown in Fig. 8, for the ZEUS measurement of dijet PHP for events with a leading neutron, the RAPGAP model that mixes standard fragmentation and pion exchange gives a good description of the shape of the $x_L$ distribution, and also shows the dominant contribution of the pion-exchange to describe the high $x_L$ region. A detailed description of the measurement can be found in [13].

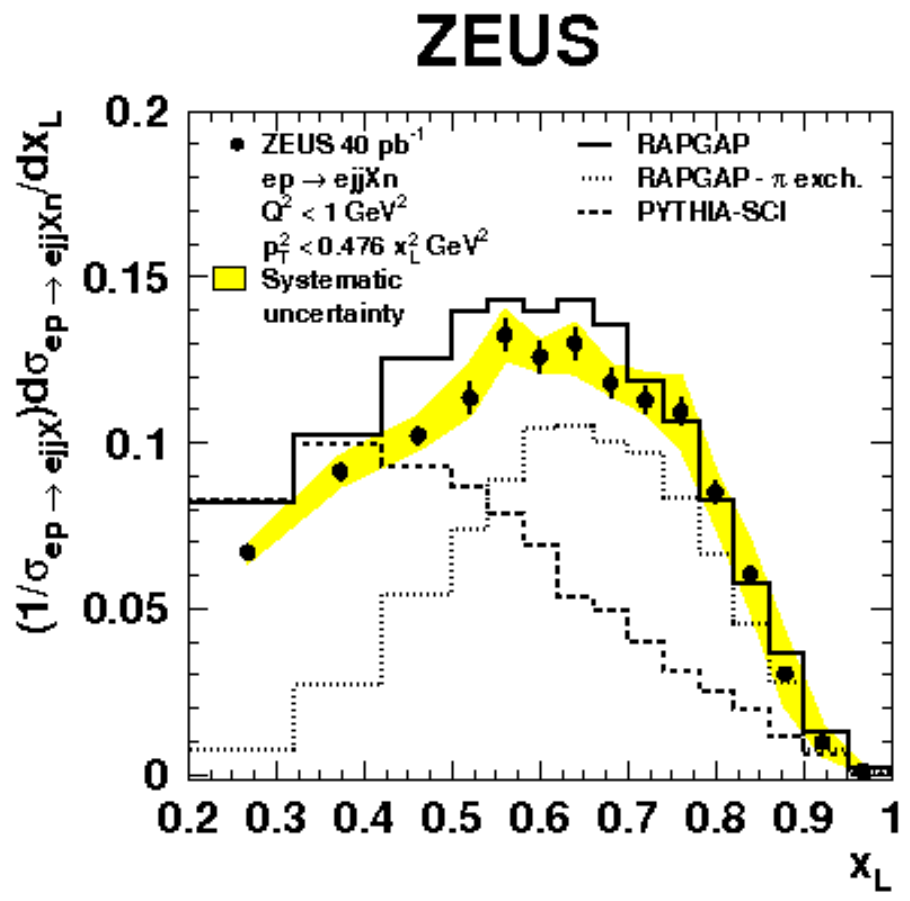

***Fig. 8:*** *The normalized differential cross section in dijet events. The solid line shows the prediction of the full RAPGAP model, the dotted line is the contribution from pion exchange and the dashed line is the prediction of PYTHIA with SCI [13].*

In the exchange picture, the cross section for a process in *ep* scattering with e.g. a LN production factorizes as

$$\sigma(ep \to enX) = f_{\pi/p}(x_L, t) \cdot \sigma(e\pi \to eX),$$

where $f_{\pi/p}$ is the flux of virtual pions in the proton, depending on $(x_L,t)$.

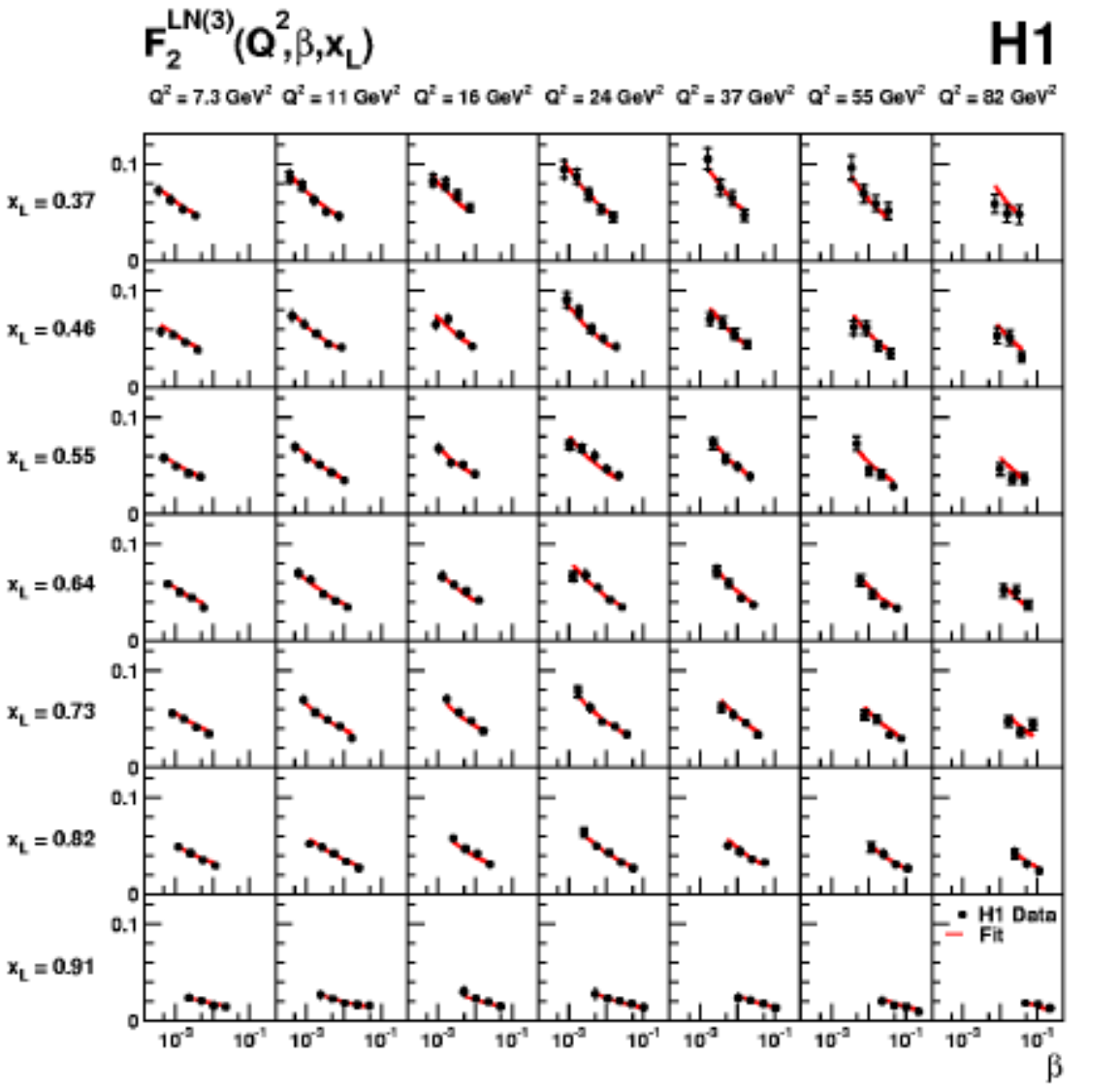

***Fig. 9:*** *The semi-inclusive structure function* $F_2^{LN(3)}$ $(Q^2,\beta,x_L)$*, for neutrons with* $p_T < 0.2$ *GeV, shown as a function of* $\beta$ *in bins of* $Q^2$ *and* $x_L$*. The lines are the results of the fit described in the* [14].

Figure 9 shows the interesting LN structure function measured recently by H1 [14] as a function of $\beta$ for different $Q^2$ and $x_L$ bins. Fitting $F_2^{LN}$ with $c(x_L)\beta^{-\lambda(Q2)}$, where $c(x_L)$ is a free normalisation parameter, we observe a $\lambda$ behaviour independent of $x_L$ (consistent with vertex factorisation) and that $\lambda$ increases with $Q^2$ from 0.23 to 0.3 (similar to the proton $F_2$) which is consistent with the hypothesis of limiting fragmentation. Assuming that the pion exchange mechanism dominates leading neutron production, the data provide constraints on the shape of the pion structure function.

## Exclusive vector mesons measurements at HERA

Diffractive PHP and electroproduction of vector mesons (VM), $\gamma^*p\rightarrow Vp$, and Deeply Virtual Compton Scattering (DVCS), $\gamma^*p\rightarrow\gamma\, p$, have been extensively studied at HERA, see for a review [1, 15,16]. They are often governed by large distance (soft) processes, however, for short distance processes, the presence of a hard scale offers the possibility to use perturbative techniques to calculate diffractive amplitudes.

The $W$ and $t$ dependence of VM cross sections, described as a soft process within the framework of Regge phenomenology, is expected to be parameterised as $W^{\delta}$ and $e^{-b|t|}$ respectively, where $b$ characterize the transverse size of the interaction.

In the presence of a hard scale, provided by large $Q^2$ values or large VM masses, $\mu^2=(Q^2+M_V^2)/4$, pQCD can be used. Furthermore, the DVCS process is dominated by two-gluons exchange with different longitudinal and transverse momenta (the skewing effect) and the measurements of the DVCS corss section at HERA provide constraints on the generalised parton distributions (GPDs).

The behaviour of $b$ and $\delta$ with increasing scale has been presented; $\delta$ rises from 0.2 (for 'soft Pomeron') to 0.8 (reflecting the strong rise at small $x \propto \mu^2/W^2$ of the gluon density in the proton), and $b$ decreases from 10 GeV$^{-2}$ to 5 GeV$^{-2}$. Figure 10 shows a collection of PHP VM cross section measurements at HERA together with other experimental results. The transition between soft and hard regime is visible by the steeper rise of the cross section going from light to heavy VM masses.

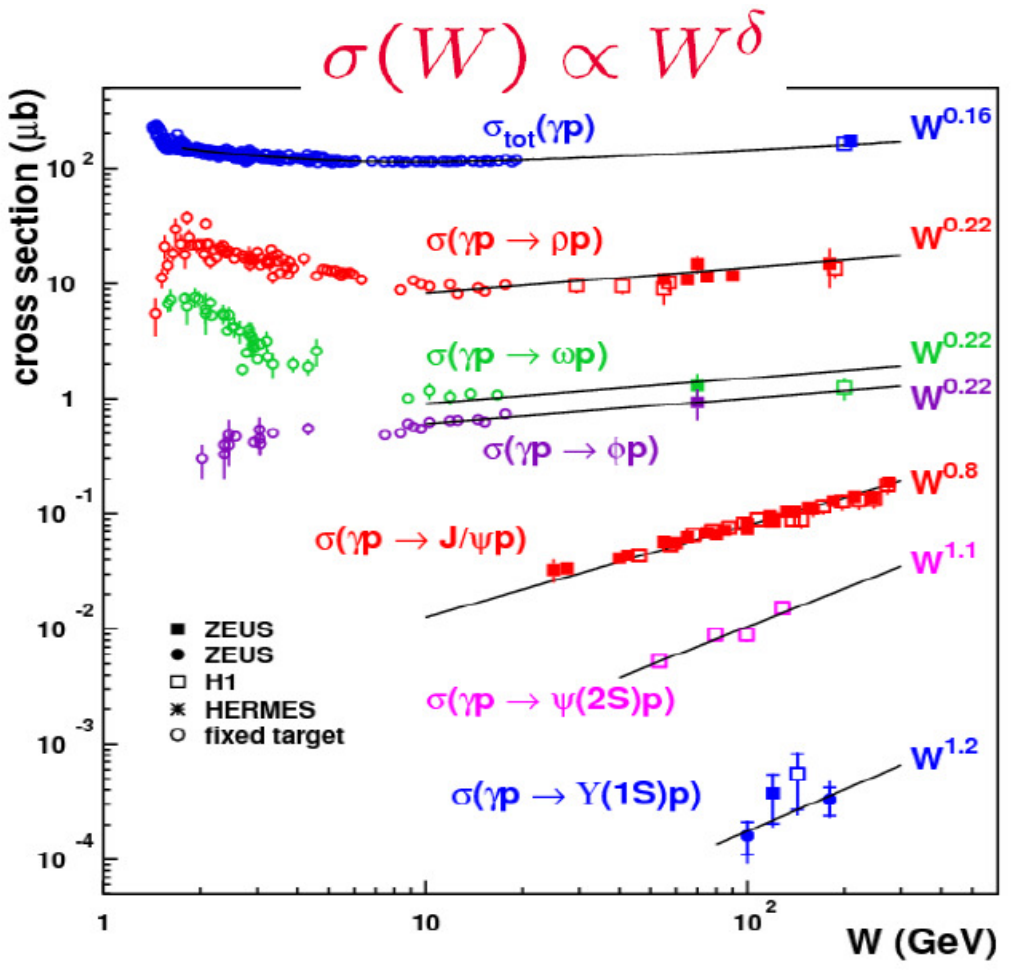

***Fig. 10:*** *Compilation of VM cross section measurements in PHP at HERA* .

A similar steeper rise with the energy $W$ is seen also in VM electroproduction and DVCS measurements when the scale $\mu^2$ increases, see Fig. 11. In this figure a different way to present the result is shown. Plotted as a function of the scale $\mu^2$ is the Pomeron intercept $a_{IP}(0)$ of the Pomeron trajectory.

The Pomeron intercept and its slope $a'_{IP}$ are related to the $\delta$ by the relation

$$\alpha_{IP}(0) = 1 + \delta / 4 + \alpha'_{IP} / \langle |t| \rangle .$$

DVCS shows a hard behaviour even at lowest $Q^2$, that may suggest that the most sensitive part to soft scale is the VM wave function.

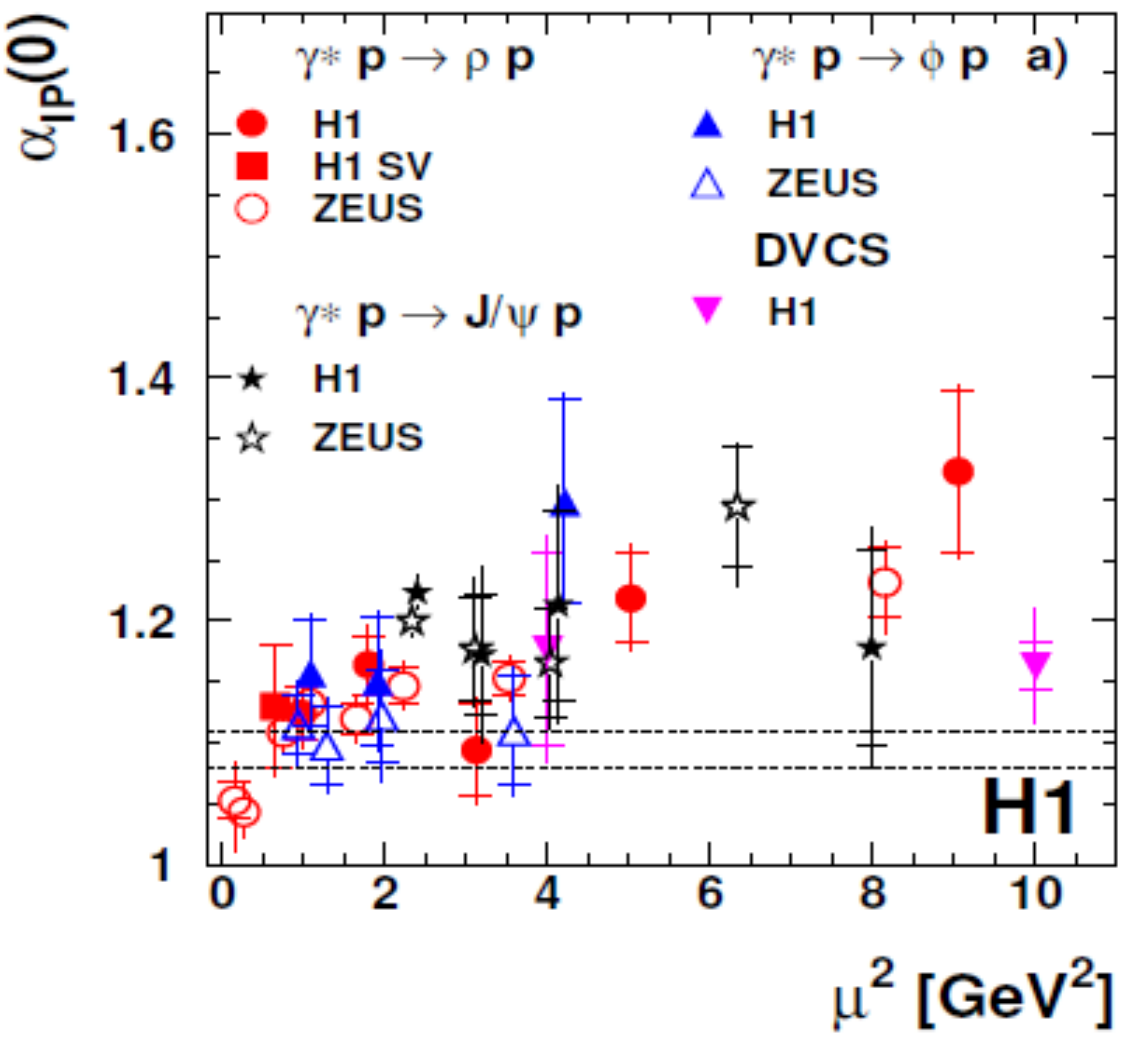

***Fig. 11:*** *Compilation of VM and DVCS $a_{IP}(0)$ as a function of the scale $\mu^2$ extracted at HERA .*

In Fig. 12 we present a compilation of the *b* slope measurements at HERA as a function of the scale $\mu^2$. A transition from the soft to the hard regime is visible for VM and DVCS. The *b* value of the DVCS differential cross section as a function of *t* obtained from ZEUS has been obtained for the first time, from a direct measurement of the t variable with the LPS (see also [16] and references therein).

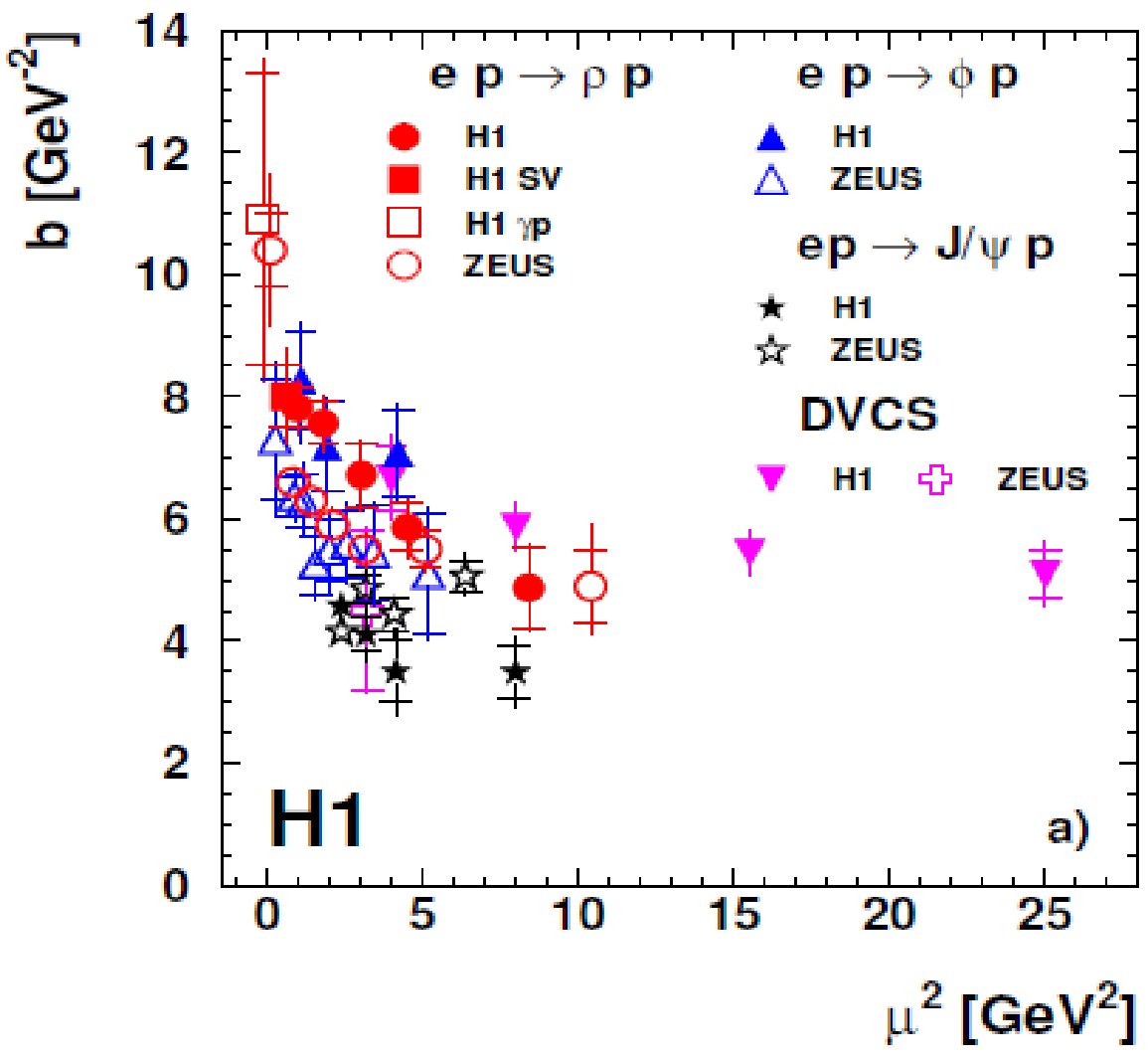

***Fig. 12:*** *Compilation of VM and DVCS b slope as a function of the scale $\mu^2$ extracted at HERA .*

## Conclusions

The HERA experiments, H1 and ZEUS, have presented new and very interesting results on diffraction.

Inclusive measurements have shown the validity of the hypothesis of vertex-factorization in diffraction and have allowed the extraction of the diffractive parton distributions. The new preliminary measurements performed by H1 in a more extended kinematic regime and the published ZEUS results will allow to further refine the results.

Additionally, the Higgs boson may be produced at the LHC via a diffractive process in which fast protons are detected. A deeper understanding of diffraction, driven by the HERA result, could therefore aid in the discovery of the Higgs boson.

Finally new measurements on Leading Baryon production and the extensive studies of vector mesons and DVCS offer the opportunity to investigate the limits of the perturbative and non-perturbative regime.

The DVCS measurement in particular are of considerable interest for the extraction of the GPDs.